\documentclass{article}  
\usepackage{bigsky2007}
\usepackage{graphicx}

\newcommand{\pt}{$p_T$}
\newcommand{\xf}{${\rm x}_F$}

\newcommand{\piplus}{$\pi^{+}$}
\newcommand{\piminus}{$\pi^{-}$}

\newcommand{\degree}{$^{\circ}$}

\frompage{000} \topage{000}                                              
\def\rightmark{BRAHMS pp results}
\def\leftmark{F.Videb\ae k}
 
\title{Results from pp at 62.4 and 200 GeV with the BRAHMS experiment}
\authors{
{F.Videb\ae k$^1$ for the BRAHMS experiment} \\[2.812mm]
{\normalsize
\hspace*{-8pt}$^1$ Brookhaven National Labloratory,
NY 11973 Upton, New York, US\\[0.2ex] 
}}
 
\abstract{Measurements of elementary pp collisions are essential for understanding heavy ion collisions. Results for pp
  collisions at 200 and 62.4 GeV are presented. At both energies NLO pQCD
  describes pion production well. The measured pion transverse single spin asymmetries
  are very large at 62.4 GeV and are reasonably well described by
  models relying on pQCD at transverse momenta larger than 1 GeV/c.}

 
\begin{document}
 
\maketitle
\setcounter{page}{1}

\section{Introduction}\label{intro}

One of the goals of the relativistic heavy ion program is to study
the properties of matter at high temperature and high density.
The explorations at the Relativistic Heavy Ion Collider (RHIC)
indicate that that at c.m. energies of 200 GeV per nucleon pair indeed
such system is formed with novel properties as characterized through
the particle production
\cite{BRAHMS:WhitePaper,PHOBOS:WhitePaper,STAR:WhitePaper,PHENIX:WhitePaper}.
Part of these conclusions rely on comparison to elementary pp
collisions, where such nuclear effects should not be present.
The RHIC experiments have also explored Au+Au and Cu+Cu collisions at 62.4
GeV, an intermediate energy between that at the SPS (18 GeV)
and that where the bulk of the RHIC heavy ion program is performed at
200 GeV.  
The 62.4 GeV was chosen to match the highest energy where data were obtained in the
ISR experiments. 
Thus,  to minimize the systematic uncertainty for the comparisons to
the heavy ion data a brief run with (polarized) pp at 62.4 GeV was
carried out to collect pp reference data by the BRAHMS, PHENIX and STAR
experiments. Thus, at this point in time an extensive set of pp reference data
have been collected at these two energies.

In addition to the pp comparison data there is an extensive spin program
at RHIC. It has long been envisioned that BRAHMS could contribute to these
studies using the transversely polarized proton beams by measuring pions
at large momenta at forward angles. 
Part of run-5 and run-6 was dedicated to such measurements. 
There is current interest in understanding the large transverse single
spin asymmetries (SSA) observed at moderate to large values of \xf\  in 
the energy range from 20 to 200 GeV in the framework of different approaches to
pQCD. 
The main theoretical efforts to account for the observed SSA have focused on the
role of transverse momentum dependent
(TMD) partonic effects in the structure of the initial transversely
polarized nucleon\cite{sivers} and the fragmentation process of a
polarized quark into hadrons\cite{collins}. Higher twist effects
(''twist-3") arising from quark-gluon correlation effects have also
been considered as a possible origin of SSA\cite{qiu,twist3review}.
To validate such an approach that is based on pQCD it is key to see 
how well pQCD describes inclusive data at high rapidities.

This contribution presents results of inclusive pp reactions at 200 and 62.4 GeV, and comparisons to pQCD at both
energies. Transverse spin results from 62.4 and 200 GeV are also described and
compared to model calculations.

\section{Inclusive pp measurements}

The data used for this analysis were collected with the BRAHMS
detector system.
The BRAHMS detector consists of two movable magnetic
spectrometers,  the Forward Spectrometer
(FS) that can be rotated from $2.3$\degree\  to $15$\degree\ , and the
Mid-Rapidity Spectrometer (MRS)
that can be rotated from $34$\degree\  to $90$\degree\  degrees relative to the beam line,
and several global detectors for measuring multiplicities and
luminosity, and  determining the interaction vertex, and
providing  a start time (T0) for time-of-flight measurement.
\begin{figure}[h]
\centering
  \includegraphics[height=3.0in]{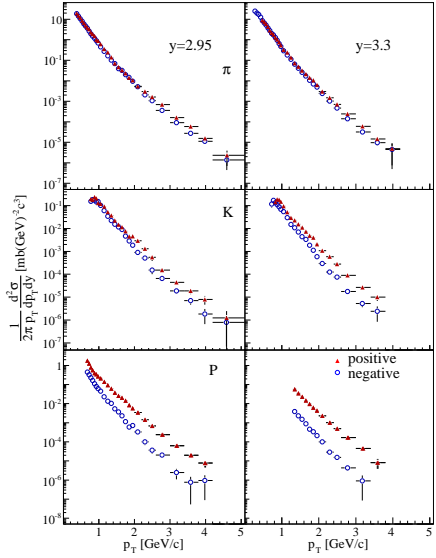}
\vspace{-0.4cm}
  \caption{\label{fig:spectra} Invariant cross section
    distributions for pion, kaons protons and anti-protons  produced in  p+p
    collisions at $\sqrt{s}=200$ GeV at rapidities
    $y=2.95$ (left panels) and  $y = 3.3$ (right panels).  
In all panels, positive charged particles are shown with  filled
triangles and negative ones with open circles. The errors displayed 
in these plots are statistical. }
\end{figure}

The MRS is a single-dipole-magnet spectrometer with a solid angle of
$\approx 5 msr$ and a magnetic bending power up to 1.2 Tm. Most of the
data presented here were recorded at  magnetic field settings of 0.4 and 0.6 Tm.
The MRS contains two time projection chambers, TPM1
and TPM2 sitting in field free regions in front of and behind the dipole (D5).
This assembly is followed by two highly segmented scintillator time-of-flight
walls, one (TOFW) at 4.51 m and a second (TFW2) at either 5.58 m
($90$\degree setting)  or 6.13 m  (other angle settings).

The FS consists of 4 dipole magnets D1, D2, D3 and
D4 with a bending power of up to 9.2 Tm. 
The spectrometer has 5 tracking stations T1 through T5, and  
particle identifiing detectors: H2, a segmented time-of-flight wall, and a  Ring Imaging Cherenkov Detector (RICH)\cite{BRAHMS:RICHNIM}.  
Additional details on the BRAHMS experimental setup can be found in
ref.~\cite{BRAHMS:NIM}.

The minimum bias trigger used to normalize these measurements is
defined with a set of Cherenkov radiators (CC) placed symmetrically 
with respect to the nominal interaction point and covering  
pseudo-rapidities that range in absolute value from 3.26 to 5.25.
This trigger required that at least one hit is detected in both sides of the array.

The present analysis was done with charged particles that originated
from  collisions of polarized protons with interaction vertices in the
range of  $\pm  40$ cm.  
For the 200 GeV data invariant cross sections were extracted 
in  narrow ($\Delta y = 0.1) $  rapidity bins centered at $y=2.95$ and $y
= 3.3$, respectively. 
Narrow rapidity bins are required to reduce the effects of rapidly
changing cross sections in particular at higher \pt. 
Each distribution is obtained from the merging of up to five magnetic field settings. 
The data are corrected for the spectrometer geometrical acceptance,
multiple scattering, weak decays and absorption in the material along
the path of the detected particles. Tracking and matching efficiencies
for each of the 5 tracking stations in the spectrometer were
calculated by constructing full tracks with only 4 stations and
evaluating the efficiency in the $5^{th}$ station. 
The overall efficiency is about 80-90\% and is included in the
extraction of the cross sections. 
Particles are identified by the RICH.
The low momentum part of the proton spectra is measured using the RICH in veto mode.

Data for $y=2.95$ and $y=3.3$ are presented for pions, kaons and protons
in Fig.\ref{fig:spectra}. The pions exhibit a power law behavior, and
at high \pt\ the ratio $\pi^{-}/\pi^{+}$ is less than one indicating
the increasing importance of the valence quarks. The $\bar{p}/p$ ratio is
much smaller than 1 \pt\ indicating that in the case of
fragmentation the gluons cannot dominate the particle production. 

\begin{figure}[ht]
\centering
\includegraphics[height=2.5in]{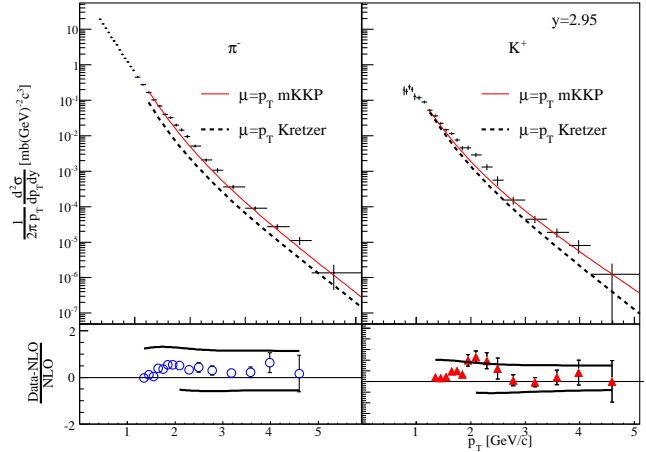}
\vspace{-0.5cm}
\caption[]{Comparison of \piminus\ and $K^{+}$ invariant spectra at
  rapidity 2.95 to NLO pQCD calculations at 200 GeV.}
\label{fig:NLO}
\end{figure}

\begin{figure}[htb]
\centering
\includegraphics[height=2.0in]{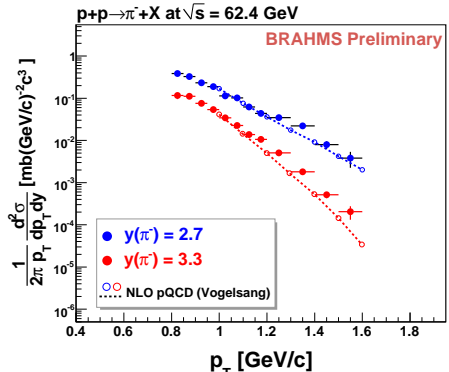}
\vspace{-0.5cm}
\caption[]{Invariant cross section for \piminus\ at rapidity 2.7 and
  3.3 at 62.4 GeV. The curves are NLO pQCD calculations as described
  in the text.}
\label{fig:NLO62}
\end{figure}

The measured differential cross-sections are compared with NLO pQCD
calculations \cite{Werner} 
evaluated at equal factorization and renormalization scales, $\mu
\equiv \mu_{F}=\mu_{R}=p_{T}$. 
These calculations use the CTEQ6 parton distribution functions
\cite{CTEQ6} 
and a modified version of the ``Kniehl-Kramer-Potter" 
(KKP) set of fragmentation functions (FFs) \cite{KKP} referred to here as mKKP, as well as the ``Kretzer" (K) set \cite{Kretzer}.  
The KKP set includes functions that fragment into  the sums
$\pi^{+}+\pi^{-}$ and  $K^{+}+K^{-}$.
Modifications were necessary  to obtain functions producing the separate charges for both $\pi $ and $K$. 
These modifications involve the following operations: to obtain the
FFs producing $\pi^{+}$, 
the functions that describe the fragmention of favored light quarks  $u, \bar{d}$ into
$\pi^{0}$ were multiplied by  $(1+z)$ ( e.g. $D_{u}^{\pi^{+}}=(
1+z)D_{u}^{\pi^{0}}$ with
$D_{u}^{\pi^{0}}=\frac{1}{2}D_{u}^{\pi^{+}+\pi^{-}}$) 
where $z$ is the fraction of the parton momentum carried by the
hadron, 
and $D_{u}^{\pi^{+}}$ the function fragmenting u quarks into positive
pions. 
The functions fragmenting  unfavored quarks $\bar{u}, d$ into
$\pi^{0}$ are multiplied by $1-z$.  
The same operation is done for  $\pi^{-}$, but this time the favored quarks are $\bar
{u}$ and $d$. 
The FFs of strange quarks and gluons  are left unmodified. 
Similar modifications were applied to obtain FFs into  $K^{+}$ and $K^{-}$,  but this time, the starting functions were the ones fragmenting $
u, \bar{u}, s, \bar{s}$ into the sum $K^{+}+K^{-}$.  
Figure \ref{fig:NLO} shows that the  agreement between the NLO
calculations that include the mKKP FFs and the measured pion cross
section is remarkable (within 20\% above 1.5 GeV/c).
Similar good agreement was obtained  for neutral pions at $y=0$ \cite{PHENIXpi0} and at $y=3.8$ \cite{STARppForward} at RHIC.
The agreement between the calculated and the measured kaon
cross-sections is equally good. 
The difference between the mKKP and Kretzer parametrizations 
is driven by higher contributions from gluons fragmenting into pions. 
This difference has been identified as an indication that the gg and gq
processes dominate the interactions at mid-rapidity \cite{PHENIXpi0}. 
The present results indicate that such continues to be the case at high
rapidity. The calculation that uses the Kretzer set underestimates the pion yields by a 
factor of $\sim 2$ at all values of \pt\  while for positive kaons
the agreement is good at low momentum but deteriorates at higher momenta.
Additional details as well as comparisons to p and $\bar{p}$ spectra
can be found in Ref.\cite{brahms:ppqcd}. 

At 62.4 GeV ,where the beam rapidity is 4.2, the spectrometer at
forward angle samples produced particles that carries
a significant fraction of the available momenta (31.2 GeV/c).
Thus particle production is influenced by the kinematic limit.
Data for identified charged hadrons were collected at
2.3\degree,3\degree, 4\degree, and 6\degree. 
Figure \ref{fig:NLO62} shows differential cross sections for \piminus\
for rapidities 2.7 and 3.3. The cross sections
changes rapidily with rapidity at high \pt\  where \xf\ values up to
0.5 are probed.
The data are compared to NLO pQCD in the same figure. The calculation
are done in the same manner as for the 200 GeV using the KPP fragmentation
function evaluated at $\mu=p_T$ scale. The calculation describes the
overall magnitude and shape quite well, though at the highest rapidity
there is a tendency for the calculation to fall below the data at the
highest \pt. This may be in agreement with the
analysis\cite{soffer:2004} of 53 GeV $\pi^{0}$ data from the ISR at
a fixed angle of 5\degree\ , comparable to the conditions for the present measurements
(2.3\degree and 4\degree), but at larger \xf\ where NLO pQCD results falls  considerably
below the data and with increasing discrepancy with \xf.
In contrast to the aforementioned paper we do, though, conclude that NLO
pQCD gives a satisfactory description of the charged pion data at high rapidity.

\section{Transverse Single Spin Asymmetries}
The SSA is defined as a ``left-right'' asymmetry of produced particles from 
the hadronic scattering of transversely polarized protons off unpolarized protons. 
Experimentally the asymmetry can be obtained 
by flipping the spins of polarized protons, and is customarily defined as 
analyzing power $A_N$: 
\begin{equation} 
A_N = \frac{1}{\mathcal P} \frac{(N^+ - {\mathcal L}N^-)}{(N^+ + {\mathcal L}N^-)}, 
\label{eq:An}
\end{equation}
where ${\mathcal P}$ is the polarization of the beam, ${\mathcal L}$ is the
spin dependent relative luminosity (${\mathcal L}$ = ${\mathcal
  L_+}$/${\mathcal L_-}$)
and $N^{+(-)}$ is the number of detected particles with beam spin
vector oriented up (down).  
\begin{figure}[htb]
\centering
\vspace{-0.1cm}
\includegraphics[height=2.0in]{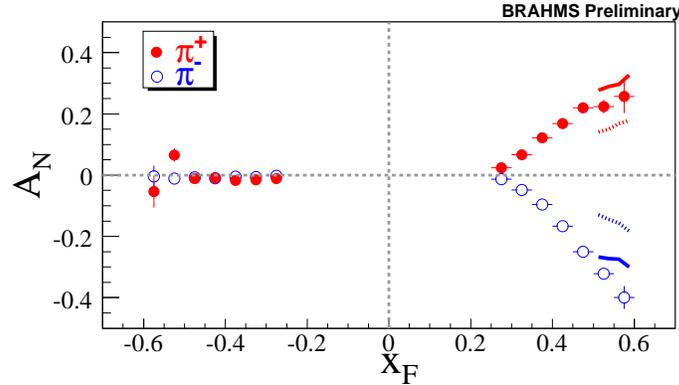}
\vspace{-0.5cm}
\caption[]{$A_N$ vs \xf\ for \piplus and \piminus\ at 62.4 GeV for
  positive and negative \xf. Solid symbols are for \piplus\ and open
  for \piminus. The curves are from the twist-3 calculations with
  (line) and without (broken) sea and anti-quark
  contribution. Predictions from the Sivers function calculation are
  shown as dotted lines.}
\label{fig:an_pi_62}
\end{figure}
The average polarization of the beam ${\mathcal P}$ as
determined from the on-line CNI measurements is about $60\%$ for
run 6 (62.4 GeV) and about $45\%$ for run 5 (200 GeV).
The systematic error on the $A_N$ measurements is estimated to be $20\%$ including
uncertainties from the beam polarization ($\sim 18\%$).  
The systematic error represents mainly scaling uncertainties on the values of $A_N$.

The production rate of pions in given \pt \ and \xf\  bins are determined in
a similar way as that for inclusive spectra, except that acceptance and
effcicieny corrections are not performed since all of these will
cancel out in the calculation of the quantities needed for
determination of $A_N$. 

The analyzing power $A_N$ for charged pions, $A_N(\pi^+)$ and 
$A_N(\pi^-)$ at  $\sqrt{s} = 62.4$ GeV as a function of $x_F$ are
shown in Fig.~\ref{fig:an_pi_62} using data taken at 2.3\degree\ and 3\degree\ .
The \pt\  range covered is from about 0.6 GeV/c at the lowest \xf\  up
to 1.5 GeV/c at the highest \xf.
The measured $A_N$ values show strong dependence with \xf\  reaching 
large asymmetries reaching up to $\sim$ 40\% at $x_F$ $\sim$ 0.6.
The $A_N$ values are positive for $\pi^+$ and negative for $\pi^-$ and
with approximately the same magnitude.
The asymmetries and their $x_F$-dependence are qualitatively in agreement
with the measurements from E704/FNAL.

\begin{figure}[htb]
\centering
\includegraphics[height=1.8in]{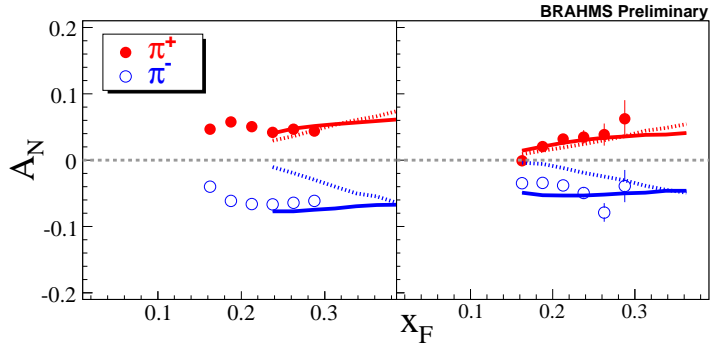}
\vspace{-0.1cm}
\caption[]{$A_N$ vs \xf\ for pions at 200 GeV.
Solid symbols are for \piplus\ and open for \piminus measured at
2.3\degree\ (left panel) and 4\degree\ (right panel). The curves are from the twist-3 calculations with
  (line) and without (broken) sea and anti-quark
  contributions. 
Predictions from the Sivers function calculation are
  shown as dotted lines.}
\label{fig:an_pi_200}
\end{figure}

The analyzing power $A_N$ for charged pions, $A_N(\pi^+)$ and 
$A_N(\pi^-)$ at  $\sqrt{s} = 200$ GeV as a function of $x_F$ are
shown in Fig.~\ref{fig:an_pi_200} using data taken at 2.3\degree\ and 4\degree\ .
The kinematic range is quite different from the 62.4 GeV data. The
\xf-range covered is from 0.15 to 0.35, while the \pt\ values covered by the
measurements are much higher ranging from $\sim 1$ to 4 GeV/c. The
left panel shows data from the 2.3\degree\ setting, while the right
panel is for 4\degree. The latter for a given \xf\ represents a higher
\pt\  value than for the 2.3\degree\ data. Overall the asymmetries are
smaller for the higher \pt's.

Figure~\ref{fig:an_pi_62} compares $A_N(\pi)$ with a pQCD calculation in 
the range of $p_T>1$ GeV/$c$ using ``extended'' twist-3 parton distributions~\cite{qiu} including 
the ``non-derivative'' contributions~\cite{feng,feng_c}. In this framework, two calculations from the model 
are compared with the data: 
two valence densities ($u_v$,$d_v$) in the ansatz with and without sea- and anti-quark contribution in the model fit.
The calculations describe the data within the uncertainties.

For the calculations shown in the figure, the dominant contributions 
to SSAs are from valence quarks, while  sea- and anti-quark contributions on SSAs are sufficiently small that 
the current measurements are not able to quantitatively constrain these contributions. 
The data are also compared with calculations using the Sivers
mechanism 
which successfully describe FNAL/E704 $A_N$ data. 
The calculations compared with the data use valence-like Sivers functions~\cite{dalesio,umberto_c} for $u$ and $d$ 
quarks with opposite sign. The fragmentation functions used are from the KKP parameterization~\cite{KKP}, but 
the Kretzer fragmentation function~\cite{Kretzer}
gives similar results.
The calculations shown with dotted lines in the figure underestimate $A_N$ for both
\pt\  ranges, which indicates that TMD parton distributions are not
sufficient to describe the SSA data at 62.4 GeV.
The same kind of calculations are compared to the 200 GeV data in
Fig.\ref{fig:an_pi_200} yielding similar conclusions as for 62.4 GeV

\section{Summary}
In summary,  unbiased invariant cross sections of identified charged
particles as function of $p_{T}$ were measured at high rapidity in p+p
collisions at $\sqrt{s}=200\ \rm{GeV}$ and  $\sqrt{s}=62.4\ \rm{GeV}$. 
NLO pQCD calculations reproduce reasonably well the produced particle
(pions and kaons) distributions. The SSA for inclusive pions were
measured at forward rapidities at the same two energies. A twist-3
pQCD model of $A_N$ describes the \xf\ and energy dependence
reasonable well. There are theoretical challenges in describing the
data at both lower energy and lower \pt.

We thank Werner Vogelsang, Feng Yuan and Umberto D'Alesio for providing us with their calculations
shown in this contribution and in forth coming publications.
This work was supported by Brookhaven Science Associates,LLC under contract DE-AC02-98-CH10886
with the U.S. Department of Energy and by a sponsered grant from Renaissance Technologies Corporation.

\vfill\eject
\end{document}